\documentstyle[amssymb,aps,prd,t1enc,amsmath,epsfig]{revtex}
\begin{document}
\draft
\title{Cyclotron damping and Faraday rotation of gravitational waves}
\author{Martin Servin$^1$, Gert Brodin$^1$ and Mattias Marklund$^{2}$}
\address{$1$ Department of Plasma Physics,\\
Ume{\aa} University, SE--901 87 Ume{\aa}, Sweden}
\address{$2$ Department of Electromagnetics,\\
Chalmers University of Technology, SE--412 96 G\"oteborg, Sweden}
\date{\today}
\maketitle

\begin{abstract}
We study the propagation of gravitational waves in a collisionless plasma
with an external magnetic field parallel to the direction of propagation.
Due to resonant interaction with the plasma particles the gravitational wave
experiences cyclotron damping or growth, the latter case being possible if
the distribution function for any of the particle species deviates from
thermodynamical equilibrium. Furthermore, we examine how the damping and
dispersion depends on temperature and on the ratio between the cyclotron-
and gravitational wave frequency. The presence of the magnetic field leads
to different dispersion relations for different polarizations, which in turn
imply Faraday rotation of gravitational waves.
\end{abstract}


% Two-column format, abstract excluded.
%\twocolumn[ 
%\hsize\textwidth\columnwidth\hsize\csname@twocolumnfalse\endcsname 

%]% Ending of two-column format

\section{Introduction}

The propagation of weak gravitational waves in the presence of matter and
electromagnetic fields has been considered by several authors, e.g. \cite
{Grishchuk}-\cite{cosmology}. The back-reaction on the gravitational waves,
resulting in damping and/or dispersion, has been studied by Refs. \cite
{Grishchuk}-\cite{Macedo}, but the gravitational effects of matter (most
frequently assumed to be a fluid or neutral gas) and the effects of
electromagnetic fields have mainly been treated separately. The interaction
of gravitational waves with a plasma -- which is the most common state of
matter relevant for gravitational wave propagation -- has been considered in
an astrophysical as well as an cosmological context, see e.g. Refs. \cite
{lin}-\cite{Ignatev97} and Ref. \cite{cosmology} respectively, and
references therein. In general it turns out that matter which is in a plasma
state -- and thus exhibits electromagnetic properties -- has possibilities
of more efficient interaction with gravitational radiation, as compared to
neutral matter.

In this paper we study the propagation of weak gravitational waves in a
collisionless plasma with an external static and homogenous magnetic field,
parallel to the direction of propagation. Naturally, the matter and fields
produces a background curvature, but nevertheless it is meaningful to treat
the background as Minkowski, provided the wavelength is much shorter than
the background curvature (see the Appendix for a detailed discussion). It
turns out that there is a new effect on the gravitational waves that appears
due to presence of the external magnetic field - a gravitational analog of 
{\it cyclotron damping} of electromagnetic waves (cf. Ref. \cite{cydamp}).%
{\bf \ }In the electromagnetic case, the waves may interact resonantly with
the gyrating motion of the particles, and the resonance occurs for particles
that experience a wave whose Doppler shifted frequency equals the
gyrofrequency. In the gravitational analog, the resonance occurs when the
(Doppler shifted) wave frequency is twice the gyrofrequency. A mechanism for
gravitational wave damping similar to cyclotron damping has been considered
previously - namely Landau damping \cite{Grishchuk}. However, cyclotron
damping is a potentially more important mechanism, since for this case,
efficient wave-particle interaction may take place without the presence of
ultra relativistic particles. The possibility for cyclotron damping of
gravitational waves has been considered by Ref. \cite{Macedo} and recently
by Ref. \cite{Kleidis}. However, the calculation of the damping coefficient
in Ref. \cite{Kleidis} do not rest on a selfconsistent kinetic theory, and
it turns out that our value of the damping based on the Vlasov equation is
smaller than theirs by several orders of magnitude. On the other hand, if we
take the limit of a non-relativistic Maxwellian distribution function, our
results essentially (see note \cite{Macedo-not}) reduces to those of Ref. 
\cite{Macedo}.

In the case where the unperturbed distribution function of the particles is
not in thermodynamical equilibrium, gravitational wave instabilities rather
than damping may occur. We give a condition on the distribution function for
instabilities to develope and demonstrate that it may be fulfilled, for
instance, by plasmas with a temperature anisotropy. Naturally, the
gravitational wave dispersion is also modified by the presence of the
magnetic field. The dependence of the damping and dispersion on the
temperature and on the cyclotron- and gravitational wave frequencies is
investigated. Furthermore, we confirm that the natural wave modes are
circularly polarized waves also in the relativistic regime and that -- as a
consequence of the different dispersion relations for these modes (in an
electron-ion type of plasma) -- gravitational waves experiences a phenomena
analogous to {\it Faraday rotation }of electromagnetic waves in such a
medium.

The paper is organized as follows: In section II we present the equations
governing our system, using a tetrad description in order to make the
interpretation of our results more straightforward. Section III reviews the
problem of single particle motion in the presence of a gravitational wave
propagating parallel to an external magnetic field. In particular it is
shown that resonant particles experiences continuous acceleration with
velocities approaching the speed of light in vacuum. In section IV the
interaction between the gravitational waves and the plasma is studied
selfconsistently in the linearized approximation using the
Einstein-Maxwell-Vlasov system of equations. The damping and dispersion of
the gravitational wave is studied in some detail for an electron-ion plasma
as well as for an electron-positron plasma in thermodynamical equilibrium.
The case of a plasma which is not in thermodynamical equilibrium is studied
in section V. Finally, in section VI, we summarize our results and discuss
their implications. There is also an appendix where , the problem of
separating effects into background curvature effects (``indirect'' matter
effect) and ``direct'' matter effects is discussed. The results in this
paper, obtained using a tetrad frame formalism, are compared with those of a
coordinate frame formalism.

\section{Basic equations}

We consider the interaction between weak gravitational waves and a
collisionless plasma in an external magnetic field. Since we consider
nonempty space the background space-time is necessarily curved. However, if
the wavelength of the gravitational waves and the interaction region is
small relative to the background curvature we may take the background to be
flat and static, and the energy-momentum tensor to be the one corresponding
to the {\it perturbations} of the electromagnetic and material fields (see
the Appendix).

For simplicity, the direction of propagation is assumed parallel to the
magnetic field which we take to be static and homogeneous. Linearized, the
Einstein field equations\ (EFE) take the form 
\begin{equation}
\square \ h_{ab}=-2\kappa \left[ \delta T_{ab}-{%
%TCIMACRO{\tfrac{1}{2}}%
%BeginExpansion
{\textstyle{1 \over 2}}%
%EndExpansion
}\delta T\eta _{ab}\right]  \label{EFE}
\end{equation}
provided the gauge condition $h_{\quad ,b}^{ab}=0$ is fulfilled, which is
equivalent to state that only tensorial perturbations are present. That the
gauge condition is indeed satisfied will be verified in Section III below.
Here $\square \ \equiv \left[ c^{-2}\partial _{t}^{2}-\partial _{z}^{2}%
\right] $, $h_{ab}$ is the small deviation from the Minkowski background
metric, i.e. $g_{ab}=\eta _{ab}+h_{ab}$, $\kappa \equiv 8\pi G/c^{4}$, $%
\delta T_{ab}$ is the part of the energy-momentum tensor containing small
electromagnetic and material field perturbations associated with the
gravitational waves and $\delta T$ $=\delta T_{\;a}^{a}$. In the following
it is understood that we neglect contributions of second order and higher in 
$h_{ab}$. In our notations $a,b,c,...=0,1,2,3$ and $i,j,k,...=1,2,3$ and the
metric has the signature $(-+++)$.

In vacuum, a linearized gravitational wave can be transformed into the
transverse and traceless (TT) gauge. Then we have the following line-element
and corresponding orthonormal frame basis 
\begin{eqnarray}
{\rm d}s^{2} &=&-c^{2}{\rm d}t^{2}+(1+h_{+}(\xi ))\,{\rm d}%
x^{2}+(1-h_{+}(\xi ))\,{\rm d}y^{2}  \nonumber \\
&&+2h_{\times }(\xi )\,{\rm d}x\,{\rm d}y+{\rm d}z^{2}\ ,  \label{le}
\end{eqnarray}
\begin{eqnarray*}
{\bf {e}}_{0} &\equiv &c^{-1}\partial _{t}\ ,\quad {\bf {e}}_{1}\equiv (1-{%
%TCIMACRO{\tfrac{1}{2}}%
%BeginExpansion
{\textstyle{1 \over 2}}%
%EndExpansion
}h_{+})\partial _{x}-{%
%TCIMACRO{\tfrac{1}{2}}%
%BeginExpansion
{\textstyle{1 \over 2}}%
%EndExpansion
}h_{\times }\partial _{y}\ , \\
{\bf {e}}_{2} &\equiv &(1+{%
%TCIMACRO{\tfrac{1}{2}}%
%BeginExpansion
{\textstyle{1 \over 2}}%
%EndExpansion
}h_{+})\partial _{y}-{%
%TCIMACRO{\tfrac{1}{2}}%
%BeginExpansion
{\textstyle{1 \over 2}}%
%EndExpansion
}h_{\times }\partial _{x}\ ,\quad {\bf {e}}_{3}\equiv \partial _{z}\ .
\end{eqnarray*}
where $\xi \equiv z-ct$ and $h_{+},h_{\times }\ll 1$. As it turns out, the
gravitational waves takes this form also in the particular case (propagation 
{\it parallel} to the magnetic field) we are considering. The difference to
the vacuum case will be that $\xi =z-v_{{\rm ph}}t$, where $v_{{\rm ph}}$ is
the phase velocity of the gravitational wave. From now on we will refer to
tetrad components rather than coordinate components.

We follow the approach presented in \cite{lin} for handling gravitational
effects on the electromagnetic and material fields. Suppose an observer
moves with 4-velocity $u^{a}$. This observer will measure the electric and
magnetic fields $E_{a}\equiv F_{ab}u^{b}$ and $B_{a}\equiv 
%TCIMACRO{\tfrac{1}{2}}%
%BeginExpansion
{\textstyle{1 \over 2}}%
%EndExpansion
\,\epsilon _{abc}F^{bc}$ , respectively, where $F_{ab}$ is the
electromagnetic field tensor and $\epsilon _{abc}$ is the volume element on
hyper-surfaces orthogonal to $u^{a}$. It is convenient to introduce a
3-vector notation ${\bf {E}}\equiv (E^{i})=(E^{1},E^{2},E^{3})$ etc. and $%
{\bf \nabla }\equiv {\bf e}_{i}$. From now on we will assume that $u^{0}=c$
is the only nonzero component of $u^{a}$. Generally, the Maxwell equations
contain terms coupling the electromagnetic field to the gravitational
radiation field. If the gravitational waves propagate parallel to a magnetic
field there are no gravitationally induced effects on ${\bf {E}}$ and ${\bf {%
B}}$. This can be deduced as follows: Given the Ricci rotation coefficients
for gravitational waves in the TT-gauge, the ``gravitational source terms''
in the Maxwell equations in Ref.\cite{lin} vanishes for the given
orientation of the magnetic field.

The equation of motion for a particle of mass $m$ and charge $q$ in an
electromagnetic and gravitational wave field is 
\begin{equation}
\frac{d}{dt}{\bf p}=q\left[ {\bf {E}}+{(}\gamma m)^{-1}{\bf p{\times B}}%
\right] -{\bf {G}}  \label{eqmot}
\end{equation}
where $\gamma =\sqrt{1+p_{i}p^{i}/(mc)^{2}}$ and the four-momenta is $%
p^{a}=\gamma mdx^{a}/dt$. The gravitational force like term $G^{i}\equiv
\Gamma _{\;ab}^{i}p^{a}p^{b}/\gamma m$, where $\Gamma _{\;ab}^{i}$ are the
Ricci rotation coefficients, becomes 
\begin{eqnarray}
G_{1} &=&{%
%TCIMACRO{\tfrac{1}{2}}%
%BeginExpansion
{\textstyle{1 \over 2}}%
%EndExpansion
}(v_{{\rm ph}}-p_{z}/\gamma m)\left[ \dot{h}_{+}p_{1}+\dot{h}_{\times }p_{2}%
\right]  \label{G1} \\
G_{2} &=&{%
%TCIMACRO{\tfrac{1}{2}}%
%BeginExpansion
{\textstyle{1 \over 2}}%
%EndExpansion
}(v_{{\rm ph}}-p_{z}/\gamma m)\left[ -\dot{h}_{+}p_{2}+\dot{h}_{\times }p_{1}%
\right]  \label{G2} \\
G_{3} &=&{%
%TCIMACRO{\tfrac{1}{2}}%
%BeginExpansion
{\textstyle{1 \over 2}}%
%EndExpansion
(}\gamma m)^{-1}\left[ \dot{h}_{+}(p_{1}^{2}-p_{2}^{2})+2\dot{h}_{\times
}p_{1}p_{2}\right]  \label{G3}
\end{eqnarray}
for weak gravitational waves propagating in the z-direction in Minkowski
space, where the dot symbol represents derivative with respect to $\xi $.

In order to account for resonant wave-particle interactions we apply kinetic
plasma theory, representing each particle species by a distribution function 
$f$ governed by the Vlasov equation. In tetrad form the Vlasov equation
reads \cite{Rendall} 
\[
{\cal L}f=0 
\]
where the Liouville operator is 
\[
{\cal L}\equiv \partial _{t}+(c/p^{0})p^{i}e_{i}+\left[ F_{{\rm EM}%
}^{i}-\Gamma _{\;ab}^{i}p^{a}p^{b}c/p^{0}\right] \partial _{p^{i}} 
\]
and the electromagnetic force responsible for geodesic deviation is $F_{{\rm %
EM}}^{i}\equiv q(E^{i}+\epsilon ^{ijk}p_{j}B_{k}/\gamma m)$. In vector
notation the Vlasov equation reads 
\begin{equation}
\partial _{t}f+\frac{{\bf p}\cdot \nabla f}{\gamma m}+\left[ q\left( {\bf E+}%
\frac{{\bf p\times B}}{\gamma m}\right) -{\bf G}\right] \cdot \nabla _{{\bf p%
}}f=0  \label{Vlasov-vec}
\end{equation}
where $\nabla _{{\bf p}}\equiv (\partial _{p_{1}},\partial _{p_{2}},\partial
_{p_{3}})$. In the absence of gravitational waves, the Vlasov equation has
the following spatially homogeneous (thermodynamical) equilibrium solution,
the Synge-J\"{u}ttner distribution\cite{Synge}, 
\begin{equation}
f_{SJ}=\frac{n_{0}\mu }{4\pi (mc)^{3}K_{2}(\mu )}e^{-\mu \gamma }  \label{SJ}
\end{equation}
where $n_{0}$ is the spatial particle number density, $\mu \equiv
mc^{2}/k_{B}T$, $k_{B}$ is the Boltzmann constant, $T$ the temperature and $%
K_{2}(\mu )$ is a modified Bessel function of second kind. Generally, the
unperturbed static solutions to Eq. (\ref{Vlasov-vec}) consistent with a
homogeneous and static magnetic field, say in the $z$-direction, but not
necessarily in thermodynamical equilibrium are distribution functions $%
f=f(p_{\perp },p_{3})$, where $p_{\perp }=\sqrt{p_{1}^{2}+p_{2}^{2}}$.

Since there are no induced electromagnetic fields, the perturbed
energy-momentum tensor can be written 
\begin{equation}
\delta T_{ab}=\sum_{{\rm p.s.}}\int \frac{p_{a}p_{b}}{m\gamma }f_{G}d^{3}p
\label{emtensor}
\end{equation}
where $f_{G}$ is the gravitational perturbation of the distribution function
and the summation is over particle species (${\rm p.s.}$). The
selfconsistent set of equations governing the interaction between the
gravitational waves and the plasma are thus the two coupled equations (\ref
{EFE}) and (\ref{Vlasov-vec}). Obviously, as we have adapted a tetrad
formalism, we mean the tetrad equivalence of Eq.(\ref{EFE}) \cite{Tetrad}.

\section{Cyclotron resonance acceleration}

As we will consider effects that are due to wave-particle interaction, it
might be in place to first review some results of single particle
(test-particle) motion in the presence of gravitational waves and an
external magnetic field.

Single particle motion in gravitational wave fields has been examined by
many authors, see e.g. \cite{Grishchuk} and references therein. Particles
moving in a monochromatic gravitational wave field experiences periodic
change in its energy and periodic deviation from its mean direction of
propagation. If the particle motion is constrained, for instance by a
magnetic field, the change in energy and momentum may be cumulative. We
refer to this as resonant acceleration.

We focus here on charged particles in a homogenous and static magnetic field
parallel to the direction of propagation of gravitational waves. The single
particle motion in this situation have been investigated in some detail by 
\cite{Kleidis}, treating it as a Hamiltonian dynamical problem, and also by 
\cite{Holten}. Most noticeable is that not only can resonant acceleration
occur, particles can even be ``trapped'' in such a resonant state and
experience essentially unlimited linear growth in kinetic energy (linear in
coordinate time) and the parallel velocity will approach the velocity of
light. The solution we present here does not -- in contrast to that of Ref. 
\cite{Kleidis} -- result in an {\it exact} description of the particle
motion but does reveal the main effect (resonant acceleration of particles)
and has the advantage of being straight forward. It also provides an
intuitively clear explanation of the mechanism and the resemblance to
electromagnetic cyclotron resonance acceleration \cite{cyclo}. In the
electromagnetic (vacuum) case this {\it phase-lock} situation also exists,
i.e., unlimited resonant acceleration of ``trapped'' particles.

Since there are no gravitationally induced electromagnetic fields (linear in 
$h_{ab}$), Eq. (\ref{eqmot}) becomes 
\begin{equation}
\frac{d}{dt}{\bf p}=(\omega _{{\rm c}}/\gamma ){\bf p{\times \hat{z}}}-{\bf {%
G\equiv F}}  \label{forceq}
\end{equation}
where $\omega _{{\rm c}}\equiv qB/m$. The gravitational wave is assumed
monochromatic and arbitrarily polarized: $h_{+}=\widehat{h}_{+}\exp [{\rm i}%
(kz-\omega t)]+{\rm c.c.}$ and $h_{\times }=\widehat{h}_{\times }\exp [{\rm i%
}(kz-\omega t)]+{\rm c.c.}$, where ${\rm c.c.}$\ stands for complex
conjugate. In this section $\omega $ and $k$ are assumed real.

We assume $\omega _{{\rm c}}/\gamma \gg \omega h_{ab}$ so that the
gravitational force in Eq. (\ref{forceq}) is small compared to the
electromagnetic one. Thus, the particle orbits are close to the gyrating
motion in the absence of a gravitational field, and we therefore make the
ansatz $p_{1}=\frac{{1}}{\sqrt{2}}\widehat{p}(t)\exp (-{\rm i}\omega _{{\rm c%
}}t/\gamma )+{\rm c.c.}$ and $p_{2}=\pm \frac{{\rm i}}{\sqrt{2}}\widehat{p}%
(t)\exp (-{\rm i}\omega _{{\rm c}}t/\gamma )+{\rm c.c.}$ where the amplitude 
$\widehat{p}(t)$ depend slowly on time [i.e.\ $|d\hat{p}/dt|\ll |(\omega _{%
{\rm c}}/\gamma )\hat{p}|$] due to the gravitational influence and $\pm =%
{\rm sgn}(q).$ Note that the cyclotron period is not itself a constant, but
is depending on the gamma factor that also is varying slowly. Consider now
the explicit form of the parallel driving force 
\begin{eqnarray}
F_{3} &=&-{%
%TCIMACRO{\tfrac{{\rm i}}{2}}%
%BeginExpansion
{\textstyle{{\rm i} \over 2}}%
%EndExpansion
}\frac{k}{\gamma m}\left[ \widehat{h}_{+}\widehat{p}^{2}\pm {\rm i}\widehat{h%
}_{\times }\widehat{p}^{2}\right] e^{{\rm i}(kz-\omega t-2\omega _{{\rm c}%
}t/\gamma )}+{\rm c.c.}  \nonumber \\
&&-{%
%TCIMACRO{\tfrac{{\rm i}}{2}}%
%BeginExpansion
{\textstyle{{\rm i} \over 2}}%
%EndExpansion
}\frac{k}{\gamma m}\left[ \widehat{h}_{+}\widehat{p}^{\ast 2}\mp {\rm i}%
\widehat{h}_{\times }\widehat{p}^{\ast 2}\right] e^{{\rm i}(kz-\omega
t+2\omega _{{\rm c}}t/\gamma )}+{\rm c.c.}  \label{driver}
\end{eqnarray}
where the star symbol denotes complex conjugate. A particle with trajectory $%
z(t)$ will typically experience an irregular oscillatory force. Unless the
particle is resonant (or almost resonant) with the wave, the parallel motion
will be random and there will be no net effect -- except for the possibility
of (small) diffusive acceleration\cite{Kleidis}. Particles may, however, be
resonant, i.e., have a trajectory such that there will be a non-oscillatory
force resulting in a lasting acceleration/deceleration over several time
periods of gyration. Almost resonant particles will be acted on by a force
varying in time on a time scale (depending on the magnitude of the mismatch)
slower than the gravitational wave period. From Eq. (\ref{driver}) we see
that there are two possibilities for a particle to be resonant and thus
acted on by a constant force: 
\begin{eqnarray*}
k\frac{{\rm d}\bar{z}}{{\rm d}t}-\omega -\frac{2\omega _{{\rm c}}}{\gamma }
&=&0 \\
k\frac{{\rm d}\bar{z}}{{\rm d}t}-\omega +\frac{2\omega _{{\rm c}}}{\gamma }
&=&0
\end{eqnarray*}
implying 
\begin{eqnarray}
\omega &=&-\frac{2\omega _{{\rm c}}}{\gamma -p_{3}/mv_{{\rm ph}}}
\label{rescond1} \\
\omega &=&\frac{2\omega _{{\rm c}}}{\gamma -p_{3}/mv_{{\rm ph}}}
\label{rescond2}
\end{eqnarray}
respectively, where $v_{{\rm ph}}=\omega /k$. By $\bar{z}$ we mean the time
averaged trajectory and since we will only consider time-averaged effects,
we have put $p_{3}=m\gamma {\rm d}\bar{z}/{\rm d}t$ here and throughout the
remainder of this section. Physically, the particles are resonant when they
see a wave whose Doppler shifted frequency is twice the gyrofrequency $%
\omega _{{\rm c}}/\gamma $. The factor two -- not present in the
electromagnetic case -- is due to the fact that the driving force is
quadratic in $\widehat{p}$. Note that the two resonance conditions can be
fulfilled {\it simultaneously} only by particles that are oppositely charged.

The question is now whether or not the resonant particles remain resonant
even though they are accelerated -- a change in $p_{3}$ and $\gamma $ may
potentially lead to a violation of the resonance condition (\ref{rescond1})
or (\ref{rescond2}). Clearly the resonance is preserved only if $\gamma
-p_{3}/mv_{{\rm ph}}$ is a constant of motion. The gravitational force, Eq. (%
\ref{G1})-(\ref{G3}), has the property ${\bf G\cdot p}=\gamma mv_{{\rm ph}%
}G_{3}$\cite{Lorentzprop}. Thus it holds that 
\begin{equation}
\frac{d}{dt}\left( \gamma -\frac{p_{3}}{mv_{{\rm ph}}}\right) =\frac{{\bf %
F\cdot p}}{\gamma m^{2}c^{2}}{\bf -}\frac{F_{3}}{mv_{{\rm ph}}}=0
\label{conserved}
\end{equation}
if and only if $v_{{\rm ph}}=\pm c$, i.e. the resonance is preserved in
vacuum but generally not in a medium. In many situations for gravitational
waves, however, the vacuum relation $v_{{\rm ph}}=\pm c$ holds to a very
good approximation.

In the remainder of this section we will confine ourself to the case $v_{%
{\rm ph}}=c$, ${\rm sgn}(q)=-1$ and to the resonance condition (\ref
{rescond2}). We define the constant of motion $\alpha \equiv $ $\gamma
-p_{3}/mc$ and observe that $p_{3}=mc(\gamma -\alpha )$ and $\left| \widehat{%
p}\right| =mc\sqrt{2\gamma \alpha -1-\alpha ^{2}}$. By considering the (time
averaged) time evolution of $\gamma $ we find 
\begin{equation}
\frac{d\gamma }{dt}=\omega \left[ 2\alpha -\gamma ^{-1}(1+\alpha ^{2})\right]
\left[ |\widehat{h}_{+}|\sin \varphi +|\widehat{h}_{\times }|\cos \psi %
\right]   \label{gamma}
\end{equation}
where $\varphi \equiv \arg (\widehat{h}_{+}\widehat{p}^{\ast 2}/\gamma m)$
and $\psi \equiv \arg (\widehat{h}_{\times }\widehat{p}^{\ast 2}/\gamma m)$.
Apparently the kinetic energy is a monotonously increasing function for
particles with suitable initial phase of the gyrating motion, i.e., for 
\[
h\equiv |\widehat{h}_{+}|\cos \varphi +|\widehat{h}_{\times }|\sin \psi >0
\]
independently of the initial magnitude of $\widehat{p}$ and $p_{3}$. From
now on we limit ourselves to such particles. For large times when $\gamma
\gg 1$, Eq. (\ref{gamma}) implies $\gamma \propto 2\alpha h\omega t$. Using
the expressions for $\widehat{p}$ and $p_{3}$ in terms of $\gamma $ and $%
\alpha $\ we note that coordinate momenta $\widehat{p}/\gamma \rightarrow 0$
as $t\rightarrow \infty $ whereas $p_{3}/\gamma $ scales as 
\begin{equation}
p_{3}/\gamma \rightarrow mc-\frac{mc}{2h\omega t}  \label{pz}
\end{equation}
This implies the possibility of unlimited acceleration in which $%
p_{3}/\gamma m$ approaches $c$ on a time scale $t\sim (h\omega )^{-1}$.
Note, however, that the assumption $\omega _{{\rm c}}/\gamma \gg \omega
h_{ab}$ formally restricts the predicting power of these results. However,
as demonstrated by Ref. \cite{Kleidis} the linear growth of $\gamma $ (due
to parallel acceleration) remain even beyond $\omega _{{\rm c}}/\gamma \gg
\omega h_{ab}$.

The sections to follow will concentrate on linearized gravitational wave
propagation, in which case the particles will be assumed to deviate only
slightly from the unperturbed orbits. It should be noted, however, that
cyclotron acceleration may have interesting applications in the vicinity of
a binary pulsars close to merging. Practically speaking the effective
distance of acceleration close to the source will be limited by effects due
to a 3-D geometry. On the other hand, particles close to pulsars are likely
to have a relativistic background temperature, in which case the resonant
ones may be accelerated to ultrahigh energies.

\section{Cyclotron damping}

As seen in the preceding section, charged particles in a homogenous static
magnetic field can be accelerated and decelerated by gravitational waves.
Thus it should not be surprising that the gravitational wave will be damped
-- or be unstable and experience growth -- as it propagates through a
collisionless plasma. The damping (or growth) rate will depend on how the
particles are distributed in momentum space. In order to incorporate the
damping effect due to this resonant wave-particle interaction mechanism, we
use a kinetic description of the plasma, i.e. each plasma component is
represented by a distribution function $f(x^{a},p^{i})$. The gravitational
waves, $h_{+}=\widehat{h}_{+}\exp [{\rm i}(kz-\omega t)]$ and $h_{\times }=%
\widehat{h}_{\times }\exp [{\rm i}(kz-\omega t)]$ are superimposed on the
Minkowski background metric (see the Appendix), and are assumed to be
associated with a small perturbation, $f_{G}=\hat{f}_{G}\exp [{\rm i}%
(kz-\omega t)]$ , of the distribution function, i.e. $f=f_{0}+f_{G}$, where $%
f_{0}$ is a stationary solution to the Vlasov equation (\ref{Vlasov-vec}) in
the absence of gravitational waves. This means that the background
distribution function $f_{0}$ is a function of $p_{\perp }$ and $p_{3}$,
where $p_{\perp }=\sqrt{p_{1}^{2}+p_{2}^{2}}$.

\subsection{Linearized Vlasov equation}

We begin by calculating the perturbed distribution function that results
from a gravitational wave propagating parallel to an external magnetic
field. Linearizing the Vlasov Eq. (\ref{Vlasov-vec}) in $h_{+}$, $h_{\times
} $ and $f_{G}$ gives 
\begin{equation}
{\rm i}\left[ k\frac{p_{3}}{\gamma m}-\omega \right] f_{G}+\frac{q}{\gamma m}%
{\bf p\times B}\cdot \nabla _{{\bf p}}f_{G}={\bf G}\cdot \nabla _{{\bf p}%
}f_{0}  \label{lin-vl}
\end{equation}
It is convenient to change to cylindrical coordinates in momentum space,
i.e. we define $p_{1}\equiv p_{\perp }\cos \phi $ , $p_{2}\equiv p_{\perp
}\sin \phi $\ and $p_{3}\equiv p_{\parallel }$ , so that $(q/\gamma m){\bf %
p\times B}\cdot \nabla _{{\bf p}}=-\gamma ^{-1}\omega _{c}\partial _{\phi }$
and Eq. (\ref{lin-vl}) thus becomes 
\begin{equation}
\left[ {\rm i}\left( \frac{kp_{\parallel }}{\gamma m}-\omega \right) -\frac{%
\omega _{c}}{\gamma }\partial _{\phi }\right] f_{G}={\rm i}\omega \left[
h_{+}\cos 2\phi +h_{\times }\sin 2\phi \right] {\cal F}_{0}  \label{fGeq}
\end{equation}
where 
\begin{equation}
{\cal F}_{0}\equiv \frac{p_{\perp }}{2v_{ph}}\left[ \left( v_{ph}-\frac{%
p_{\parallel }}{\gamma m}\right) \partial _{p_{\perp }}f_{0}+\frac{p_{\perp }%
}{\gamma m}\partial _{p_{\parallel }}f_{0}\right]  \label{F0}
\end{equation}
Eq. (\ref{fGeq}) has the solution 
\begin{eqnarray}
f_{G} &=&\frac{1}{2}\frac{\gamma \omega {\cal F}_{0}e^{{\rm i}2\phi }}{%
kp_{\parallel }/m-\gamma \omega -2\omega _{c}}(h_{+}-{\rm i}h_{\times }) 
\nonumber \\
&&+\frac{1}{2}\frac{\gamma \omega {\cal F}_{0}e^{-{\rm i}2\phi }}{%
kp_{\parallel }/m-\gamma \omega +2\omega _{c}}(h_{+}+{\rm i}h_{\times })
\label{fG}
\end{eqnarray}
The occurrence of singularities in Eq. (\ref{fG}) indicate that there is a
resonant interaction. In the case $\omega _{c}=0$ there are no
singularities, because without magnetic field $\omega \geq ck$ in our
approximation, and therefore Landau damping of gravitational waves do not
occur. For any finite value of $\omega _{c}$, however, the expression for $%
f_{G}$ has singularities and we therefore expect cyclotron damping to occur.

\subsection{Dispersion relation}

We now solve the tetrad equivalence \cite{Tetrad} of EFE (\ref{EFE}). Using (%
\ref{emtensor}) and (\ref{fG}) it is straight forward to confirm that the
TT-gauge is a consistent choice in our case. The two diagonal elements both
reads

\begin{eqnarray}
\left[ \omega ^{2}-c^{2}k^{2}\right] h_{+} &=&2c^{2}\kappa \sum_{{\rm p.s.}%
}\int\limits_{-\infty }^{\infty }\int\limits_{0}^{2\pi
}\int\limits_{0}^{\infty }\frac{p_{\perp }^{3}}{\gamma m}\cos ^{2}\phi
f_{G}dp_{\perp }d\phi dp_{\parallel }  \nonumber \\
&=&\frac{1}{2}(h_{+}-{\rm i}h_{\times })I_{-}+\frac{1}{2}(h_{+}+{\rm i}%
h_{\times })I_{+}  \label{hp}
\end{eqnarray}
and the two off diagonal elements gives

\begin{eqnarray}
\left[ \omega ^{2}-c^{2}k^{2}\right] h_{\times } &=&2c^{2}\kappa \sum_{{\rm %
p.s.}}\int\limits_{-\infty }^{\infty }\int\limits_{0}^{2\pi
}\int\limits_{0}^{\infty }\frac{p_{\perp }^{3}}{\gamma m}\cos \phi \sin \phi
f_{G}dp_{\perp }d\phi dp_{\parallel }  \nonumber \\
&=&\frac{{\rm i}}{2}(h_{+}-{\rm i}h_{\times })I_{-}-\frac{{\rm i}}{2}(h_{+}+%
{\rm i}h_{\times })I_{+}  \label{hc}
\end{eqnarray}
where we have made use of Eq. (\ref{fG}) and 
\begin{equation}
I_{\pm }=\sum_{{\rm p.s.}}C\int\limits_{-\infty }^{\infty
}\int\limits_{0}^{\infty }p_{\perp }^{3}\frac{{\cal F}_{0}}{kp_{\parallel
}/m-\gamma \omega \mp 2\omega _{c}}dp_{\perp }dp_{\parallel }  \label{Idef}
\end{equation}
together with $C\equiv \pi \omega \kappa c^{2}/m$. The two equations (\ref
{hp}) and (\ref{hc}) combines to 
\begin{eqnarray}
\left[ \omega ^{2}-c^{2}k^{2}-A_{+}-{\rm i}B_{+}\right] (h_{+}+{\rm i}%
h_{\times }) &=&0  \label{pmode} \\
\left[ \omega ^{2}-c^{2}k^{2}-A_{-}-{\rm i}B_{-}\right] (h_{+}-{\rm i}%
h_{\times }) &=&0  \label{mmode}
\end{eqnarray}
where 
\[
A_{\pm }\equiv 
%TCIMACRO{\func{Re}}%
%BeginExpansion
\mathop{\rm Re}%
%EndExpansion
I_{\pm }\qquad \text{and}\qquad B_{\pm }\equiv 
%TCIMACRO{\func{Im}}%
%BeginExpansion
\mathop{\rm Im}%
%EndExpansion
I_{\pm } 
\]
The natural gravitational wave modes for EFE with the given source are thus
the circularly polarized modes $h_{+}+{\rm i}h_{\times }$ and $h_{+}-{\rm i}%
h_{\times }$. Eqs. (\ref{pmode}) and (\ref{mmode}) contain the information
about the dispersion and the damping or growth (due to $B_{\pm }\neq 0$).
Due to the smallness of the gravitational coupling constant, $A_{\pm }$ and $%
B_{\pm }$ can typically be considered small in the sense that the dispersion
relations Eq. (\ref{pmode}) and (\ref{mmode}) reads 
\begin{equation}
\omega \approx ck+A_{\pm }/2\omega +{\rm i}B_{\pm }/2\omega
\label{Dr-approx}
\end{equation}
Throughout the remainder of this paper we will have this approximation in
mind and we will occasionally make use of $\omega \approx ck$ to simplify
the expressions for $I_{\pm }$.

The fact that $A_{+}\neq A_{-}$ , unless the plasma components are of equal
particle masses and as long $\omega _{{\rm c}}\neq 0$, implies that the two
gravitational wave modes have slightly different phase velocities. Thus, an
incident linearly polarized wave -- being a superposition of the two
circularly polarized states -- will experience a polarization shift, i.e.
the direction of linear polarization will be rotated as it propagates
through the medium. In the case of electromagnetic waves this is known as
Faraday rotation. In principle this could be an important result as the
polarization of a gravitational wave carries valuable information about the
emitting source \cite{gwastronomy}, e.g. the inclination of the spin axis of
a quadrupole moment source.

The pole contribution in the integral in (\ref{Idef}) is dealt with in the
standard fashion, i.e. by letting the contour of integration pass below the
pole. It should be noted that this approach disregards what happens with \
the distribution function close to the singularity, and it also misses some
other aspects of the damping process, see e.g.\cite{newLandau}, but it is
the simplest way to find the main effect due to the pole, and for our
purposes it suffices.

\subsection{Equilibrium plasma}

For a plasma in a state of thermodynamical equilibrium the unperturbed
relativistic expression for the distribution function is the
Synge-J\"{u}ttner distribution $f_0=f_{SJ}$, defined in Eq. (\ref{SJ}), and
for this choice ${\cal F}_0$ reduces to 
\begin{equation}
{\cal F}_0=-\frac{\mu p_{\perp }^2}{2\gamma (mc)^2}f_{SJ}  \label{FSJ}
\end{equation}
As no instabilities can develope the gravitational waves will exhibit
damping. In this section we examine how the damping and dispersion depends
on the ratio $\omega _{{\rm c}}/\omega $ and on the temperature of the
plasma. First we examine the nonrelativistic regime for which analytic
results can be obtained. In the relativistic regime we then present results
for an electron-positron plasma and an electron-ion plasma obtained by
numerical integration of Eq. (\ref{Idef}). For this purpose it is practical
to introduce normalized momenta ${\frak p}_{\perp }\equiv p_{\perp }/mc$, $%
{\frak p}_{\parallel }\equiv p_{\parallel }/mc$ and a dimensionless
frequency ratio $\Omega =\omega _{{\rm c}}/\omega $.

\subsubsection{Nonrelativistic temperature}

In the regime of nonrelativistic particle velocities Eq. (\ref{Idef})
together with (\ref{FSJ}) becomes 
\begin{equation}
I_{\pm }=-\sum_{{\rm p.s.}}\frac{32\pi }{\tau ^{2}}\left( \frac{\mu }{2\pi }%
\right) ^{3/2}\mu ^{-2}\int\limits_{-\infty }^{\infty }\frac{e^{-\mu {\frak p%
}_{\parallel }^{2}/2}}{{\frak p}_{\parallel }-1\mp 2\Omega }d{\frak p}%
_{\parallel }  \label{Inr}
\end{equation}
after performing the $p_{\perp }$ integration, where we have introduced $%
\tau \equiv 1/\sqrt{\pi mn_{0}G}$, which is the characteristic time for
gravitational contraction of a gas with density $n_{0}$ and particle mass $m$%
. In the case of a cosmological plasma $\tau $ coincides, apart from a
factor $\sqrt{3/8}$, with the Hubble time. From Eq. (\ref{Inr}) we obtain to
lowest order in the temperature 
\begin{eqnarray}
A_{\pm } &=&\sum_{{\rm p.s.}}\frac{16}{\tau ^{2}\mu }\frac{1}{1\pm 2\Omega }
\label{Anr} \\
B_{\pm } &=&-\sum_{{\rm p.s.}}\frac{8\sqrt{2\pi }}{\tau ^{2}\sqrt{\mu }}%
e^{-\mu (1\mp 2\Omega )^{2}/2}  \label{Bnr}
\end{eqnarray}
Note that for Eq. (\ref{Anr}) to apply we must assume $\mu ^{-1}(1\pm
2\Omega )^{-1}\ll 1$, i.e. for the wave frequency close enough to twice the
gyrofrequency, higher order thermal effects is always important for the wave
dispersion, whereas Eq.(\ref{Bnr}) -- which follows from the residue theorem
-- holds for all values of $\Omega $. Still, it is clear that the dispersion
can be enhanced by the magnetic field due to the existence of the resonance.
The exponential decrease of the damping with $1\mp 2\Omega $ implies that
significant damping only occurs in a limited region in frequency space close
to $1\mp 2\Omega =0$ for a low temperature plasma. In the limit of zero
temperature the size of this region tends to zero. The magnitude of both $%
A_{\pm }$ and $B_{\pm }$ are monotonically increasing with temperature and
vanishes at zero temperature. The results in this subsection essentially
agrees with Ref. \cite{Macedo}, see note \cite{Macedo-not}. In the case of
no magnetic field, $\Omega =0$, the dispersion relations (\ref{pmode}) and (%
\ref{mmode}) both reduces to $\omega ^{2}-c^{2}k^{2}-A=0$, where $A\equiv
\sum_{{\rm p.s.}}16/\tau ^{2}\mu $. This expression is in agreement \cite
{Numerical note} with Refs. \cite{Grishchuk}- \cite{Polnarev}.

\subsubsection{Electron-positron plasma}

Denote $\tau _{{\rm ep}}\equiv 1/\sqrt{\pi m_{{\rm e}}n_{0}G}$ ({\rm e} and 
{\rm p}\ stands for electron and positron, respectively). In this case only
the sign of the charge differ for the two particle species and we may write
Eq. (\ref{Idef}), applying (\ref{FSJ}), as 
\begin{equation}
I_{\pm }(\tilde{\Omega})=-\tau _{{\rm ep}}^{-2}\sum_{{\rm p.s.}}\frac{\mu
^{2}}{K_{2}(\mu )}\int\limits_{-\infty }^{\infty }\int\limits_{0}^{\infty }%
\frac{{\frak p}_{\perp }^{5}}{\sqrt{1+{\frak p}_{\perp }^{2}+{\frak p}%
_{\parallel }^{2}}}\frac{e^{-\mu \sqrt{1+{\frak p}_{\perp }^{2}+{\frak p}%
_{\parallel }^{2}}}}{{\frak p}_{\parallel }-\sqrt{1+{\frak p}_{\perp }^{2}+%
{\frak p}_{\parallel }^{2}}\mp 2{\rm sgn}(q)\tilde{\Omega}}d{\frak p}_{\perp
}d{\frak p}_{\parallel }
\end{equation}
where $\tilde{\Omega}=\tilde{\omega}_{c}/\omega $ and $\tilde{\omega}%
_{c}=|q|B/m$. Note that $I_{\pm }(\tilde{\Omega})=I_{\mp }(\tilde{\Omega}%
)=I_{\pm }(-\tilde{\Omega})$. The normalized function ${\cal I}_{\pm }(%
\tilde{\Omega})\equiv I_{\pm }\mu \tau _{{\rm ep}}^{2}$ is numerically
calculated for four different temperatures, namely $\mu =100$
(``nonrelativistic'') $10$, $1$ and $0.1$ (``ultrarelativistic''). The
results are displayed in Figure 1 (note that ${\cal I}_{\pm }$ is normalized
against the temperature). In the nonrelativistic case (see Figure 1. (a)) we
have a finite region, $-1/2\lesssim \tilde{\Omega}\lesssim 1/2$ , where $%
%TCIMACRO{\func{Re}}%
%BeginExpansion
\mathop{\rm Re}%
%EndExpansion
I_{\pm }$ is positive. $%
%TCIMACRO{\func{Re}}%
%BeginExpansion
\mathop{\rm Re}%
%EndExpansion
I_{\pm }$ changes sign at $\tilde{\Omega}\approx \pm 1/2$ and approaches
zero like $1/(\mp \tilde{\Omega})$ in the limit $\tilde{\Omega}\rightarrow
\pm \infty $. $%
%TCIMACRO{\func{Im}}%
%BeginExpansion
\mathop{\rm Im}%
%EndExpansion
I_{\pm }$ is negative definite (and thus there is indeed {\it damping} of
the gravitational wave for any finite $\tilde{\Omega}$) and has a gaussian
shape around the resonances $\tilde{\Omega}\approx \pm 1/2$.

This is also the characteristic behavior for $%
%TCIMACRO{\func{Re}}%
%BeginExpansion
\mathop{\rm Re}%
%EndExpansion
I_{\pm }$ and $%
%TCIMACRO{\func{Im}}%
%BeginExpansion
\mathop{\rm Im}%
%EndExpansion
I_{\pm }$ at higher temperatures. The relativistic effects on the dispersion
and the damping are the following: {\it i)} The resonance peaks of $%
%TCIMACRO{\func{Im}}%
%BeginExpansion
\mathop{\rm Im}%
%EndExpansion
I_{\pm }$ (which occur at $\tilde{\Omega}=\pm 1/2$ at zero temperature) are
shifted to higher values of $|\tilde{\Omega}|$ for higher temperatures. Also 
$%
%TCIMACRO{\func{Re}}%
%BeginExpansion
\mathop{\rm Re}%
%EndExpansion
I_{\pm }$ experiences a similar shift. In the ultrarelativistic case (see
Figure 1. (d)), $%
%TCIMACRO{\func{Re}}%
%BeginExpansion
\mathop{\rm Re}%
%EndExpansion
I_{\pm }$ changes sign at $\tilde{\Omega}\approx \pm 25$ and $%
%TCIMACRO{\func{Im}}%
%BeginExpansion
\mathop{\rm Im}%
%EndExpansion
I_{\pm }$ is centered about $\tilde{\Omega}\approx \pm 20$. {\it ii)} The
magnitude of both $%
%TCIMACRO{\func{Re}}%
%BeginExpansion
\mathop{\rm Re}%
%EndExpansion
I_{\pm }$ and $%
%TCIMACRO{\func{Im}}%
%BeginExpansion
\mathop{\rm Im}%
%EndExpansion
I_{\pm }$ {\it increases} roughly linearly with temperature (note that the
curves are normalized against $(\mu \tau _{{\rm ep}}^{2})^{-1}$ in Figure
1.), which can be compared with the temperature dependence in the
nonrelativistic regime that is given by Eq. (\ref{Anr}) and Eq. (\ref{Bnr}). 
{\it iii)} The region of damping {\it broadens}. In the cold limit $%
%TCIMACRO{\func{Im}}%
%BeginExpansion
\mathop{\rm Im}%
%EndExpansion
I_{\pm }$ takes the form of two separated gaussian functions, the width
tending to zero with diminishing temperature. These regions are widened and
the gaussian shape is deformed with increasing temperature. In the
ultrarelativistic case $%
%TCIMACRO{\func{Im}}%
%BeginExpansion
\mathop{\rm Im}%
%EndExpansion
I_{\pm }$ decays exponentially as $\tilde{\Omega}\rightarrow \pm \infty $
but approaches zero more abruptly as $\tilde{\Omega}\rightarrow 0$.

\subsubsection{Electron-ion plasma}

Denote $\tau =\tau _{{\rm ei}}\equiv 1/\sqrt{\pi m_{{\rm i}}n_0G}$ ({\rm e}
and {\rm i}\ stands for electron and ion, respectively). For an electron-ion
plasma there is an asymmetry between the particle species due to the small
mass ratio $\varepsilon \equiv m_{{\rm e}}/m_{{\rm i}}$, giving different
order of magnitudes for the two cyclotron frequencies $\omega _{c{\rm e}}$
and $\omega _{c{\rm i}}$. Thus the resonances will occur for very different
gravitational wave frequencies. Given Eq. (\ref{FSJ}), we have the following
form of Eq. (\ref{Idef})

\begin{eqnarray}
I_{\pm } &=&-\tau _{{\rm ei}}^{-2}\frac{\varepsilon \mu _{{\rm e}}^{2}}{%
K_{2}(\mu _{{\rm e}})}\int\limits_{-\infty }^{\infty
}\int\limits_{0}^{\infty }\frac{{\frak p}_{\perp }^{5}}{\sqrt{1+{\frak p}%
_{\perp }^{2}+{\frak p}_{\parallel }^{2}}}\frac{e^{-\mu _{{\rm e}}\sqrt{1+%
{\frak p}_{\perp }^{2}+{\frak p}_{\parallel }^{2}}}}{{\frak p}_{\parallel }-%
\sqrt{1+{\frak p}_{\perp }^{2}+{\frak p}_{\parallel }^{2}}\pm 2\tilde{\Omega}%
_{{\rm e}}}d{\frak p}_{\perp }d{\frak p}_{\parallel }  \nonumber \\
&&-\tau _{{\rm ei}}^{-2}\frac{\mu _{{\rm i}}^{2}}{K_{2}(\mu _{{\rm i}})}%
\int\limits_{-\infty }^{\infty }\int\limits_{0}^{\infty }\frac{{\frak p}%
_{\perp }^{5}}{\sqrt{1+{\frak p}_{\perp }^{2}+{\frak p}_{\parallel }^{2}}}%
\frac{e^{-\mu _{{\rm i}}\sqrt{1+{\frak p}_{\perp }^{2}+{\frak p}_{\parallel
}^{2}}}}{{\frak p}_{\parallel }-\sqrt{1+{\frak p}_{\perp }^{2}+{\frak p}%
_{\parallel }^{2}}\mp 2\tilde{\Omega}_{{\rm i}}}d{\frak p}_{\perp }d{\frak p}%
_{\parallel }
\end{eqnarray}
where 
\[
\mu _{{\rm e/i}}=\frac{m_{{\rm e/i}}c^{2}}{k_{B}T_{{\rm e/i}}}\qquad \text{%
and}\quad \tilde{\Omega}_{{\rm e/i}}\equiv \frac{\tilde{\omega}_{c{\rm e/i}}%
}{\omega }=\frac{|q|B}{m_{{\rm e/i}}\omega }
\]
Provided that $T_{{\rm e}}=T_{{\rm i}}$, $\mu _{{\rm i}}$ and $\mu _{{\rm e}}
$ differs typically by three orders in magnitude. This implies that the ions
can be considered nonrelativistic even in the regime of ultrarelativistic
electrons ($\mu _{{\rm e}}\sim 0.1$). It is convenient to consider the two
frequency domains $\tilde{\Omega}_{{\rm e}}\sim 1$ and $\tilde{\Omega}_{{\rm %
i}}\sim 1$ separately. Noting that $\tilde{\Omega}_{{\rm i}}=\varepsilon 
\tilde{\Omega}_{{\rm e}}$ and $\mu _{{\rm i}}=\varepsilon ^{-1}\mu _{{\rm e}}
$, in the region $\tilde{\Omega}_{{\rm e}}\sim 1$ we have 
\begin{equation}
I_{\pm }(\tilde{\Omega}_{{\rm e}})\approx -\tau _{{\rm ep}}^{-2}\frac{\mu _{%
{\rm e}}^{2}}{K_{2}(\mu _{{\rm e}})}\int\limits_{-\infty }^{\infty
}\int\limits_{0}^{\infty }\frac{{\frak p}_{\perp }^{5}}{\sqrt{1+{\frak p}%
_{\perp }^{2}+{\frak p}_{\parallel }^{2}}}\frac{e^{-\mu _{{\rm e}}\sqrt{1+%
{\frak p}_{\perp }^{2}+{\frak p}_{\parallel }^{2}}}}{{\frak p}_{\parallel }-%
\sqrt{1+{\frak p}_{\perp }^{2}+{\frak p}_{\parallel }^{2}}\pm 2\tilde{\Omega}%
_{{\rm e}}}d{\frak p}_{\perp }d{\frak p}_{\parallel }+\frac{16}{\tau _{{\rm %
ep}}^{2}\mu _{{\rm e}}}\frac{1}{1\pm 2\varepsilon \tilde{\Omega}_{{\rm e}}}
\end{equation}
The normalized function ${\cal I}_{+}(\tilde{\Omega}_{{\rm e}})\equiv
I_{+}\mu \tau _{{\rm ep}}^{2}$ is displayed in Figure 2 for $\mu _{{\rm e}%
}=100$ (``nonrelativistic'') $10$, $1$ and $0.1$ (``ultrarelativistic''). By
symmetry, $I_{-}(\tilde{\Omega}_{{\rm e}})$ is the mirror image of $I_{+}(%
\tilde{\Omega}_{{\rm e}})$, i.e. $I_{-}(\tilde{\Omega}_{{\rm e}})=I_{+}(-%
\tilde{\Omega}_{{\rm e}})$ and therefore this curve is not presented. Except
for the lack of symmetry about $\tilde{\Omega}_{{\rm e}}=0$ the result is
similar to that of the electron-positron plasma. In the given temperature
and frequency domains the ion contribution to the normalized function $%
I_{\pm }(\tilde{\Omega}_{{\rm e}})$ is just the approximately constant value 
$16$.

The effect of resonant ions becomes important in the frequency domain $%
\tilde{\Omega}_{{\rm i}}\sim 1$. In Fig. 3 we show ${\cal I}_{+}(\tilde{%
\Omega}_{{\rm i}})\equiv I_{+}\mu \tau _{{\rm ep}}^{2}$ for $\mu _{{\rm e}%
}=0.1$. The corresponding figures are qualitatively similar in the entire
temperature domain $0.1-100$, however, and therefore only one of them is
shown. The small region near $\tilde{\Omega}_{{\rm i}}=0$ ($\tilde{\Omega}_{%
{\rm i}}\sim \varepsilon $ to be specific) that contains the electron
contribution has been left out.\newline

\subsubsection{The group velocity}

The fact that $A_{\pm }$ is at some points negative implies, together with (%
\ref{Dr-approx}), that for some wavelengths and frequencies the group
velocity of the waves exceeds the speed of light. Superluminal group
velocities for gravitational waves have been found before, see Ref. \cite
{Peters} and references therein. In most cases, but not in all, it has been
an effect of the background curvature. It should be noted that several
results from the literature are in contradiction with each other. Naturally,
in our case the superluminal group velocity is a direct effect of the
medium. For the case of an electron-positron plasma the group velocity
corresponding to Eq. (\ref{Dr-approx}) is 
\begin{equation}
v_{g}\equiv \frac{d\omega }{dk}=c\left[ 1-\frac{1}{2\omega ^{2}}\left(
A_{\pm }+\frac{\tilde{\omega}_{{\rm c}}}{\omega }A_{\pm }^{\prime }\right) %
\right]
\end{equation}
where the prime denotes derivative with respect to $\tilde{\omega}_{{\rm c}%
}/\omega $. For $\omega _{{\rm c}}=0$, i.e. in the case of no magnetic
field, the medium is just a collisionless gas of charged particles and the
group velocity is smaller than the velocity of light, in agreement with the
results of previous authors, e.g. \cite{Grishchuk}. From Figure 1 it is
clear that there are regions where $v_{g}>c$ is realized, for instance about
the point $\tilde{\omega}_{{\rm c}}/\omega \approx 0.6$ (where $A_{\pm }$
has a local minima) in Figure 1(a). Similarly, also an electron-ion plasma
allows superluminal group velocities. Group velocities that exceeds the
speed of light is not necessarily at odds with causality -- an issue
explored for instance in Ref. \cite{Peters}. The group velocity can simply
not be interpreted as the (gravitational wave) signal velocity in this
situation.

\section{Nonequilibrium plasma}

In the case of thermodynamical nonequilibrium, the system has free energy
that may feed a gravitational wave instability. This occurs whenever the
imaginary part of $I_{\pm }$, defined by Eq. (\ref{Idef}), is somewhere
positive. Applying the residue theorem, Eq. (\ref{Idef}) gives

\begin{equation}
B_{\pm }=\pi \sum_{{\rm p.s.}}C\int\limits_{0}^{\infty }p_{\perp }^{3}\left[ 
\frac{{\cal F}_{0}}{\Delta _{{\rm r}}k/m}\right] _{p_{\parallel }=p_{{\rm r}%
}}dp_{\perp }  \label{Bcondrel}
\end{equation}
where $p_{{\rm r}}\equiv mc\left( \gamma _{{\rm r}}\pm 2\omega _{c}/\omega
\right) $ is the resonant parallel momenta, 
\[
\gamma _{{\rm r}}\equiv \frac{2\omega _{c}}{\omega ^{2}-c^{2}k^{2}}\left\{
\omega -ck\left[ 1-\frac{(\omega ^{2}-c^{2}k^{2})}{4\omega _{c}^{2}}\left( 1+%
\frac{p_{\perp }^{2}}{m^{2}c^{2}}\right) \right] ^{1/2}\right\} 
\]
is the gamma factor evaluated at the resonant momenta and 
\[
\Delta _{{\rm r}}\equiv 1-\frac{\omega }{ck}\frac{p_{{\rm r}}}{\gamma _{{\rm %
r}}mc} 
\]
Instabilities occurs for distribution functions and frequencies such that
the condition $B_{\pm }>0$ is satisfied. For simplicity we focus on
nonrelativistic temperatures from now on, in which case Eq. (\ref{Bcondrel})
reduces to

\begin{equation}
B_{\pm }=\sum_{{\rm p.s.}}\frac{4\pi ^{2}}{n_{0}(mc)^{2}\tau ^{2}}%
\int\limits_{0}^{\infty }p_{\perp }^{3}\left[ 4p_{{\rm c}}f_{0}+p_{\perp
}^{2}\partial _{p_{\parallel }}f_{0}\right] _{p_{\parallel }=p_{{\rm r}%
}}dp_{\perp }  \label{Bcond}
\end{equation}
where $p_{{\rm c}}\equiv \pm 2mc\omega _{c}/\omega $. In order to show that
there indeed exist instabilities we consider the following example of a
temperature-anisotropic ``drifting Maxwellian'' distribution function 
\begin{equation}
f_{{\rm drift}}=\frac{n_{0}}{\pi ^{3/2}p_{{\rm th}\parallel }p_{{\rm th}%
\perp }^{2}}e^{-[(p_{\parallel }-p_{{\rm d}})^{2}/p_{{\rm th}\parallel
}^{2}+p_{\perp }^{2}/p_{{\rm th}\perp }^{2}]}  \label{fdrift}
\end{equation}
where $p_{{\rm th}\parallel }\equiv \sqrt{2mk_{B}T_{\parallel }}$, $p_{{\rm %
th}\perp }\equiv \sqrt{2mk_{B}T_{\perp }}$ and $p_{{\rm d}}$\ is the drift
momenta, for which $B_{\pm }$ becomes 
\begin{equation}
B_{\pm }=\sum_{{\rm p.s.}}\frac{8\sqrt{\pi }}{\tau ^{2}}\frac{p_{{\rm th}%
\perp }^{2}}{mcp_{{\rm th}\parallel }}\left[ \frac{p_{{\rm d}}}{mc}-1-\alpha
\Theta \right] e^{-m^{2}c^{2}\Theta ^{2}/p_{{\rm th}\parallel }^{2}}
\label{Bex}
\end{equation}
where $\alpha \equiv T_{\perp }/T_{\parallel }-1$ and $\Theta \equiv 1\pm
2\omega _{c}/\omega -p_{{\rm d}}/mc$. It is easily seen that $B_{\pm }>0$
for certain values of $\alpha $ and $\omega _{c}/\omega $. Note that if $%
\alpha =0$ then $B_{\pm }<0$ for all values of $\omega _{c}/\omega $ and
hence there can be no temperature-{\em isotropic} beam instability \cite
{Beam}. On the other hand, putting $p_{{\rm d}}=0$, it is clear that a
temperature-anisotropic distribution function without a drift can be the
source of an instability.

The gravitational waves produced by a homogeneous plasma, due to the above
cyclotron resonance instability will have a frequency of the order of the
cyclotron frequency, at least in the nonrelativistic temperature limit
considered here. In principle this opens up the possibility of emission of
high frequency gravitational waves through large magnetic fields (of the
order of $10^{-6}$ ${\rm T}$ or larger), i.e. frequencies well above the
frequency range expected from ``conventional'' gravitational wave sources,
such as compact binaries, neutron-star normal modes and gravitationally
collapsing objects, reaching up to $10$ ${\rm kHz}$. Still the radiation
considered here is generated by a collective process where the amount of
matter interacting {\it coherently} can be as large as in other
astrophysical examples.

It is not so easy to find astrophysical applications of the cyclotron
resonance instability, however: Firstly there must be a magnetized plasma
cloud, with a nonequilibrium distribution function that fulfills the
condition $B_{\pm }>0$. Such a cloud could in principle be generated, for
example if there is a magnetic field geometry that allows for a loss-cone
distribution to evolve. However, there is an obvious risk that there will be
purely electromagnetic instabilities, which typically have much higher
growth rates than the gravitational one, that will dominate the picture.
Secondly, for significant gravitational generation, the plasma cloud must be
very much denser than the average density of the universe, otherwise the
growth of the amplitude will take place slowly even compared to a
cosmological time-scale. In a dense plasma cloud, on the other hand,
gravitation must be balanced by pressure gradients in order not to
selfcontract before significant radiation generation due to the cyclotron
resonance instability occurs. This suggests that maybe plasma
inhomogeneities should be included in our treatment, although it seems
likely that the effects of inhomogeneity may be neglected as long as the
gradient scale lengths is much longer than the wavelength of the
gravitational radiation.

\section{Summary and discussion}

We have considered linearized gravitational waves in the short wavelength
approximation, propagating in a plasma parallel to an external magnetic
field. In vacuum, there is the possibility of cyclotron acceleration of
charged particles up to velocities arbitrarily close to the speed of light.
Taking the collective effects of particle distributions on the gravitational
waves into account, it follows that the (gravitational) normal modes for the
system are circularly polarized gravitational waves, and we derive the
corresponding dispersion relations, which coincide with that of Ref. \cite
{Macedo} in the limit of a low-temperature Maxwellian plasma. In the case of
an equilibrium plasma the waves are shown to be damped due to resonant
interaction with the plasma particles and the dispersion is modified and
enhanced as compared to the case of no magnetic field. In the case of
thermodynamical nonequilibrium, there is the possibility of gravitational
wave instabilities. To show in a concrete way that this can be realized, we
demonstrate that there are temperature anisotropic distribution functions
that are unstable. Furthermore, we have examined how the damping and
dispersion in an electron-positron type of plasma and an electron-ion
plasma, respectively, depends on the ratio $\omega _{c}/\omega $ and on $T$
in the regime of relativistic temperature. The strongest effects occurs when 
$\omega $ and $\omega _{{\rm c}}$ are comparable and the effect increases
with temperature and density.

The question is whether cyclotron damping can be observed, if we assume that
gravitational wave astronomy \cite{gwastronomy} develops successfully. The
calculations made in Ref. \cite{Kleidis} end up with an estimate of 10\%
damping during a propagation distance of the order of 30 kpc, which suggests
that there is at least some chance of gravitational cyclotron damping to be
observed. Combining Eq. (\ref{Dr-approx}) with the definition ${\cal I}_{\pm
}\equiv I_{\pm }\mu \tau ^{2}$ (recall that ${\cal I}_{\pm }$ is the
normalized value of $I_{\pm }$ displayed in Figure1, 2 and 3) we note that
our damping rate is 
\begin{equation}
\Gamma _{\pm }=\frac{\mathop{\rm Im}{\cal I}_{\pm }}{2\omega \mu \tau ^{2}}.
\label{imomega}
\end{equation}
Similarly, the time-scale for gravitational wave dispersion is $\mu \omega
\tau ^{2}/%
%TCIMACRO{\func{Re}}%
%BeginExpansion
\mathop{\rm Re}%
%EndExpansion
{\cal I}_{\pm }$. For typical values for (equilibrium) plasma in
interstellar space, it is clear that the damping rate predicted by Eq. (\ref
{imomega}) is several orders of magnitude smaller than the estimation made
in Ref.\cite{Kleidis}. Presumably the discrepancy is due to the fact that in
Ref.\cite{Kleidis}, only the effect of {\it acceleration} of particles
(corresponding to energy being transported from the gravitational wave) is
considered. Generally when the damping due to wave-particle interaction is
small, it should be noted that as $\omega t_{{\rm prop}}$(where $t_{{\rm prop%
}}$ is the time of wave- particle interaction) grows, the energy loss of the
decelerated particles becomes very close to the energy gained by the
accelerated particles, and thus the omission of the contribution from
decelerated particles leads to large errors in the damping coefficient{\it . 
}If the singularities is treated properly, the effect of decelerating as
well as accelerating particles on the wave is included automatically in a
kinetic description. In general for parallel propagation we find that
cyclotron damping as well as dispersion of gravitational waves {\it through
interstellar }space is negligible, in the sense that there is essentially no
hope of detecting it with gravitational wave detectors with realistic
sensitivity.

In general gravitational waves propagate in an angle to the magnetic field.
It is well-known (see e.g. \cite{Grishchuk}) that this leads to generation
of electromagnetic fields by the gravitational wave, and a dispersion
relation governing propagation perpendicular to a magnetic field in a plasma
has been derived by Ref. \cite{Macedo}. The results shows that the
gravitational wave is most affected by the matter when the frequency matches
one of the natural frequencies of the system, like the cyclotron frequency
or the plasma frequency. It should be noted that in an inhomogeneous medium
with slowly varying background parameters we will typically reach a point
where the generated electromagnetic fields fulfills the dispersion relation
of some natural plasma mode. In that case linear mode conversion, which is a
resonant process involving {\it all particles, }may take place{\it .}
However, such a problem remains an issue for future research.

\section{Appendix}

At a first sight our results seem to disagree with those of Ref. \cite
{Macedo}. However, we will show below that our results essentially are in
agreement\cite{Macedo-not}, although the comparison is non-trivial. In
general the effect of a curved background space-time cannot be neglected.
However, for linearized gravitational waves with short wavelength compared
to the background curvature, the back-reaction on the gravitational wave can
be separated into two effects, those respective contribution can be added to
the flat vacuum dispersion relation $\omega ^{2}-k^{2}c^{2}=0,$ see Ref. 1
p. 427. The first effect (that is considered by us) is a direct consequence
of matter and fields (where the background curvature can be neglected), and
the second effect is an indirect consequence due to the background curvature
produced by the matter and fields . How to find the curved background
contribution to the gravitational wave dispersion relation for a
Robertson-Walker or a Schwarzschild background metric, see Ref.\cite
{Grishchuk} and Ref. \cite{Madore} respectively. Here a warning is strongly
motivated, however. In general the separation of contributions to the
dispersion relation into the ``direct'' and ``indirect'' effect is {\it not
completely unique}, but to some extent depends on the formalism used. Thus
great care must be taken when adding a ``direct'' and an ``indirect''
contribution to the dispersion relation derived by different authors, to see
that the formalisms that have produced the different expressions are
compatible.

To illustrate the above matters we compare our calculations using a tetrad
frame formalism with those of a coordinate frame formalism. In the absence
of gravitational wave perturbations the Einstein field equation reads 
\begin{equation}
R_{\mu \nu }^{(0)}=\kappa \lbrack T_{\mu \nu }^{(0)}-{%
%TCIMACRO{\tfrac{1}{2}}%
%BeginExpansion
{\textstyle{1 \over 2}}%
%EndExpansion
}g_{\mu \nu }^{(0)}T^{(0)}]
\end{equation}
In this section we use Greek indices $\mu ,\nu ,...=0,1,2,3$ (or $t,x,y,z$)
for coordinate components and reserve $a,b,...=0,1,2,3$ for tetrad
components. Adding a small perturbation, so that $R_{\mu \nu }=R_{\mu \nu
}^{(0)}+\delta R_{\mu \nu }$, $T_{\mu \nu }=T_{\mu \nu }^{(0)}+\delta T_{\mu
\nu }$ and $g_{\mu \nu }=g_{\mu \nu }^{(0)}+h_{\mu \nu }$, gives the
linearized equation 
\begin{equation}
\delta R_{\mu \nu }=\kappa \delta \lbrack T_{\mu \nu }-{%
%TCIMACRO{\tfrac{1}{2}}%
%BeginExpansion
{\textstyle{1 \over 2}}%
%EndExpansion
}g_{\mu \nu }T]
\end{equation}
Focusing on the direct effect of matter and assuming the short wavelength
regime we use $g_{\mu \nu }^{(0)}=\eta _{\mu \nu }$, and after suitable
gauge transformations we obtain: 
\begin{equation}
\square \ h_{\mu \nu }=-2\kappa \left[ \delta T_{\mu \nu }-{%
%TCIMACRO{\tfrac{1}{2}}%
%BeginExpansion
{\textstyle{1 \over 2}}%
%EndExpansion
}\delta T\eta _{\mu \nu }\right]   \label{waveq}
\end{equation}
It should be noted that in principle the left hand side of Eq. (\ref{waveq})
should contain ``cross terms'' proportional to the product of $h_{\mu \nu }$
and components of the background part of the Riemann tensor (see Ref. \cite
{Grishchuk}), corresponding to the ``indirect effects of matter'' but these
terms are omitted, not because they are small, but because in the short
wavelength limit their contribution can be calculated separately and added
afterwards. For the same reason also a term proportional to $T^{(0)}h_{\mu
\nu }=-R^{(0)}h_{\mu \nu }$\ have been omitted from the right hand side. Eq.
(\ref{waveq}) also applies to the tetrad description if it is understood
that $h_{ab}$ denotes what $\delta R_{ab}$ reduces to in this approximation,
i.e. $h_{11}=-h_{22}=h_{+}$ and $h_{12}=h_{21}=h_{\times }$ in the TT-gauge.
The short wavelength approximation and TT-gauge is now assumed throughout
the remainder of this section if nothing else is said.

The total energy-momentum tensor due to electromagnetic fields and matter is 
\[
T_{\mu \nu }=\mu _{0}^{-1}\left( F_{\mu }^{\;\sigma }F_{\nu \sigma }-\frac{1%
}{4}g_{\mu \nu }F^{\sigma \tau }F_{\sigma \tau }\right) +\sum_{{\rm p.s.}%
}\int p_{\mu }p_{\nu }f\frac{\sqrt{|g|}}{m\gamma }d^{3}p 
\]
where $\mu _{0}$ is the magnetic permeability, $\gamma \equiv p_{t}/mc=\left[
1+{\bf p}^{2}/(mc)^{2}-h^{\mu \nu }p_{\mu }p_{\nu }/(mc)^{2}\right] ^{1/2}$
and $p_{t}$\ is the zero component of the four-momenta satisfying $p^{\mu
}p_{\mu }=-m^{2}c^{2}$. Note that the definition $\gamma \equiv p_{t}/mc$\
differs from the tetrad formalism. The background magnetic field is taken to
be $F_{12}=F^{12}=B$ (being {\it identical} to that in the tetrad
description). Hence the linearized energy-momentum tensor reads 
\begin{eqnarray}
\delta T_{\mu \nu } &=&\mu _{0}^{-1}\left( -h^{\sigma \tau }F_{\sigma \mu
}F_{\tau \nu }-\frac{1}{2}h_{\mu \nu }B^{2}\right)  \nonumber \\
&&+\sum_{{\rm p.s.}}\int p_{\mu }p_{\nu }\frac{f_{0}\delta \gamma }{m\gamma
_{0}}d^{3}p+\sum_{{\rm p.s.}}\int p_{\mu }p_{\nu }\frac{\delta f}{m\gamma
_{0}}d^{3}p  \label{dT}
\end{eqnarray}
where $\delta \gamma =-\frac{1}{2}h^{\mu \nu }p_{\mu }p_{\nu }/\left( \gamma
_{0}mc\right) ^{2}$ and $\gamma _{0}=\left[ 1+{\bf p}^{2}/(mc)^{2}\right]
^{1/2}$. Note that only the last term on the right hand side appears in the
tetrad equation corresponding to Eq. (\ref{dT}), since we have $g_{ab}=\eta
_{ab}$ in that case. In some sense the perturbation of the electromagnetic
part of the energy momentum tensor is somewhat artificial, since the
electromagnetic field itself is not perturbed, although it is clear from a
technical point of view that all terms in Eq. (\ref{dT}) must be included.

In the coordinate description the Vlasov equation reads 
\[
\partial _{t}f+\frac{{\bf p}}{m\gamma }\cdot {\bf \nabla }f+\left[ {\bf F}-%
{\bf G}\right] \cdot {\bf \nabla }_{{\bf p}}f=0 
\]
where $F^{i}=qp^{\mu }F_{\;\mu }^{i}/\gamma m$, $G^{i}=\Gamma _{\mu \nu
}^{i}p^{\mu }p^{\nu }/\gamma m$, $\Gamma _{\mu \nu }^{i}$ are the
Christoffel symbols corresponding to $g_{\mu \nu }=\eta _{\mu \nu }+h_{\mu
\nu }$ and ${\bf \nabla =(}\partial _{x},\partial _{y},\partial _{z}{\bf )}$%
. The linearized Vlasov equation for the gravitationally perturbed part of
the distribution function, $f_{G}$, thus becomes 
\begin{equation}
\left[ \partial _{t}+\frac{{\bf p}}{m\gamma _{0}}\cdot {\bf \nabla }+{\bf F}%
_{0}\cdot {\bf \nabla }_{{\bf p}}\right] f_{G}+\left[ \delta {\bf F}-{\bf G}%
\right] \cdot {\bf \nabla }_{{\bf p}}f_{0}=0  \label{dVlasov}
\end{equation}
where 
\begin{eqnarray}
\delta {\bf F\cdot \nabla }_{{\bf p}}f_{0} &=&2\omega _{c}\gamma _{0}^{-1}%
\left[ h_{+}\sin 2\phi -h_{\times }\cos 2\phi \right] {\cal F}_{0}
\label{dF} \\
{\bf G\cdot \nabla }_{{\bf p}}f_{0} &=&i\left( 2\omega -kp{_{\parallel }(}%
\gamma _{0}m)^{-1}\right) \left[ h_{+}\cos 2\phi -h_{\times }\sin 2\phi %
\right] {\cal F}_{0}  \label{G}
\end{eqnarray}
and ${\cal F}_{0}$ was defined in Eq. (\ref{F0}). The term $\delta {\bf F}$
in Eq. (\ref{dVlasov}) has its origin from the lowering of an index on $%
F_{\;\mu }^{i}$ and therefore does not occur in the tetrad description \cite
{Macedo-not}. Also the term ${\bf G\cdot \nabla }_{{\bf p}}f_{0}$ differs
from the tetrad description because $G_{1}=\frac{1}{2}G_{x}$, $G_{2}=\frac{1%
}{2}G_{y}$, and $G_{3}=G_{z}$.

Solving Eq. (\ref{dVlasov}) and applying the solution to Eq. (\ref{waveq})
gives the dispersion relations 
\begin{eqnarray}
\left[ \omega ^{2}-c^{2}k^{2}-A_{+}-{\rm i}B_{+}\right] (h_{+}+{\rm i}%
h_{\times }) &=&0  \label{dr-coord-1} \\
\left[ \omega ^{2}-c^{2}k^{2}-A_{-}-{\rm i}B_{-}\right] (h_{+}-{\rm i}%
h_{\times }) &=&0  \label{dr-coord-2}
\end{eqnarray}
where $A_{\pm }=%
%TCIMACRO{\func{Re}}%
%BeginExpansion
\mathop{\rm Re}%
%EndExpansion
I_{\pm }+A_{0}$, $B_{\pm }=%
%TCIMACRO{\func{Im}}%
%BeginExpansion
\mathop{\rm Im}%
%EndExpansion
I_{\pm }$ and 
\begin{eqnarray}
I_{\pm } &=&\frac{\pi \kappa c^{2}}{m}\sum_{{\rm p.s.}}\int\limits_{-\infty
}^{\infty }\int\limits_{0}^{\infty }p_{\perp }^{3}\frac{2\omega -kp{%
_{\parallel }/(}\gamma _{0}m)\pm 2\omega _{c}/\gamma _{0}}{kp_{\parallel
}/m-\gamma _{0}\omega \mp 2\omega _{c}}{\cal F}_{0}dp_{\perp }dp_{\parallel }
\label{cI} \\
A_{0} &=&\kappa c^{2}\mu _{0}^{-1}B^{2}+\frac{\pi \kappa }{2}\sum_{{\rm p.s.}%
}\int\limits_{-\infty }^{\infty }\int\limits_{0}^{\infty }\frac{p_{\perp
}^{5}}{(\gamma _{0}m)^{3}}f_{0}dp_{\parallel }dp_{\perp }  \label{A0}
\end{eqnarray}
A couple of things should be noted. Firstly, the coordinate frame
description gives a term contained in $A_{0}$ proportional to $B^{2}$ that
remains even in the absence of particles. This term has no correspondence in
the tetrad formalism, and the reason is that there are no perturbations of
the electromagnetic part of the energy momentum tensor when we study the
projections on the tetrad basis vectors, in contrast to what happens when we
look at the coordinate basis components (cf. Eqs. (\ref{emtensor}) and (\ref
{dT})). However, the division into the perturbed and unperturbed {\it %
tensors }are the same independent of formalism, and thus the ``direct term''
proportional to $B^{2}$ in the coordinate frame formalism is compensated by
a term associated with the background curvature produced by the unperturbed
magnetic field in the tetrad formalism. On the other hand the division into
a direct effect of matter and an indirect (background curvature) effect is
not completely artificial: The direct effect is determined {\it entirely} by
the local matter content, whereas the background curvature effect is
determined by the matter content both locally and globally. Furthermore, the
background curvature effect will necessarily give contributions to (\ref
{dr-coord-1}) and (\ref{dr-coord-2}) that are real and {\it independent} of
the gravitational wave frequency (cf. the discussion after Eq. (\ref{waveq}%
)). Noting that 
\begin{equation}
\frac{2\omega -kp{_{\parallel }/(}\gamma _{0}m)\pm 2\omega _{c}/\gamma _{0}}{%
kp_{\parallel }/m-\gamma _{0}\omega \mp 2\omega _{c}}=\frac{\omega }{%
kp_{\parallel }/m-\gamma _{0}\omega \mp 2\omega _{c}}-\gamma _{0}
\label{rewrite}
\end{equation}
it is clear that the correction to the vacuum dispersion relations in the
coordinate and tetrad frame formalism deviates only by a frequency
independent real constant, and in particular the cyclotron damping agrees
perfectly within the short wavelength approximation scheme. Adding the
contribution to $I_{\pm }$ from the second term in Eq. (\ref{rewrite}) to $%
A_{0}$ then all differences between the tetrad and coordinate frame
formalisms are collected in this term: 
\begin{equation}
\kappa c^{2}\mu _{0}^{-1}B^{2}+\pi \kappa \sum_{{\rm p.s.}%
}\int\limits_{-\infty }^{\infty }\int\limits_{0}^{\infty }\left[ \frac{%
p_{\perp }^{5}}{2(\gamma _{0}m)^{3}}f_{0}-\gamma _{0}p_{\perp }^{3}{\cal F}%
_{0}c^{2}/m\right] dp_{\parallel }dp_{\perp }  \label{difference}
\end{equation}
Relaxing the assumption of Minkowskian background, i.e. adding ``cross
terms'' (cf. Ref. \cite{Grishchuk}) to Eq. (\ref{waveq}) (which of course
requires a different tetrad basis in that formalism), we could in principle
confirm the agreement of the total dispersions relations including both
direct and indirect effects. This would be a tedious task, however, since we
must then solve for the background configuration including the anisotropic
magnetic field contribution. Since we have shown that the separation into
background curvature and ``direct'' matter effects is not unique, one can
question the relevance of the figures showing the {\it real }(dispersive)
part of $I_{\pm }$ since only the direct matter effect is contained (recall
that the damping contribution is unique, however, and due to the direct
effect only). But it turns out that there are three reasons that make also
the real plots of \ $I_{\pm }$ relevant: Firstly, the main contribution to
the real value of \ $I_{\pm }$ occurs in the frequency regime when cyclotron
resonance effects plays a role, which always is a direct effect of matter 
{\it independent of formalism}. Secondly, for many cases the dominant
contribution to the background curvature comes from the rest mass ($T_{00}$)
part of the energy momentum tensor, and if that is the case, the difference
displayed in (\ref{difference}) is negligible as compared to other
background curvature contributions \cite{basvektor not}. For example we
could add the background curvature contributions for a Robertson-Walker or a
Schwarzschild background metric, using the results of Ref.\cite{Grishchuk}
and Ref. \cite{Madore} respectively, since the difference in formalisms used
in most cases would give only a negligible contribution in comparison with
that included. Thirdly, since the background curvature gives only a real
constant contribution to the dispersion relations, our plots always reveal
the frequency dependence of the dispersion.

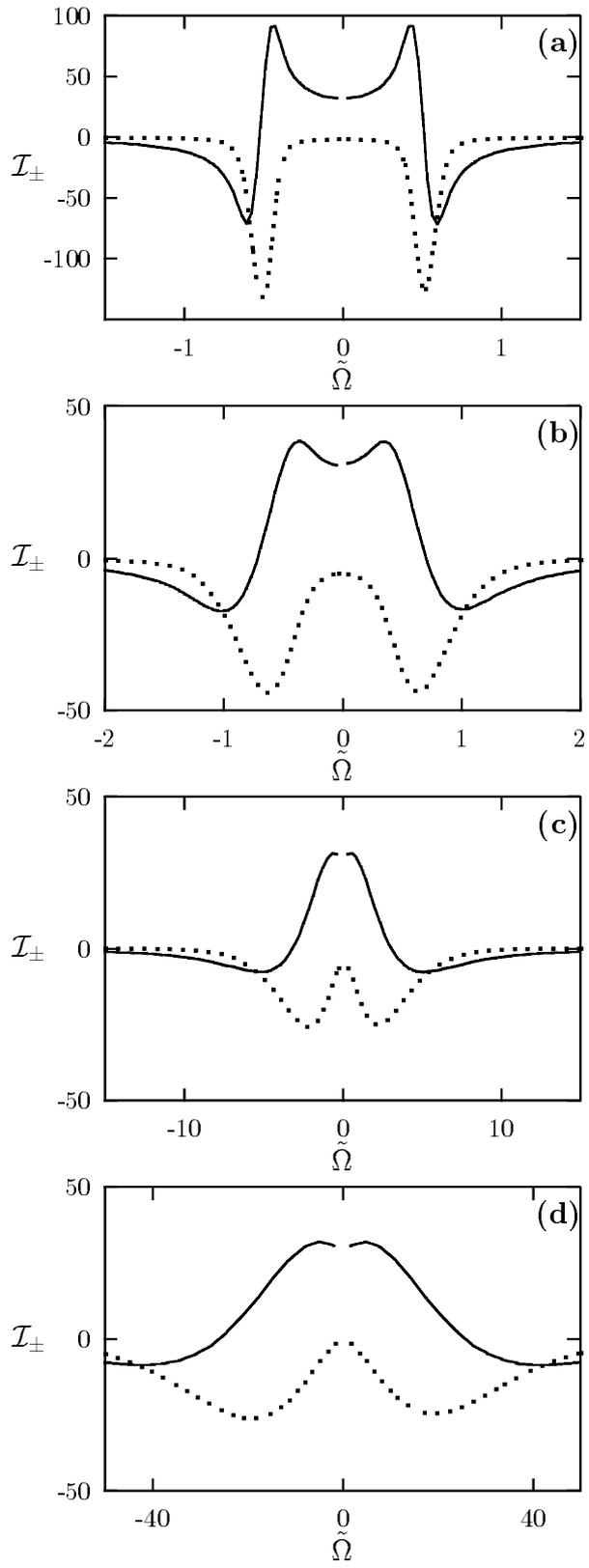
{\small FIG. 1. The real (solid line) and imaginary (dotted line) part of }$%
{\cal I}_{\pm }\equiv I_{\pm }\mu \tau _{{\rm ep}}^{2}$ {\small for an
electron-positron plasma at four different temperatures: (a) }$\mu =100$%
{\small , (b) }$\mu =10${\small \ ,(c) }$\mu =1${\small \ and (d) }$\mu =0.1$%
{\small .}\newline

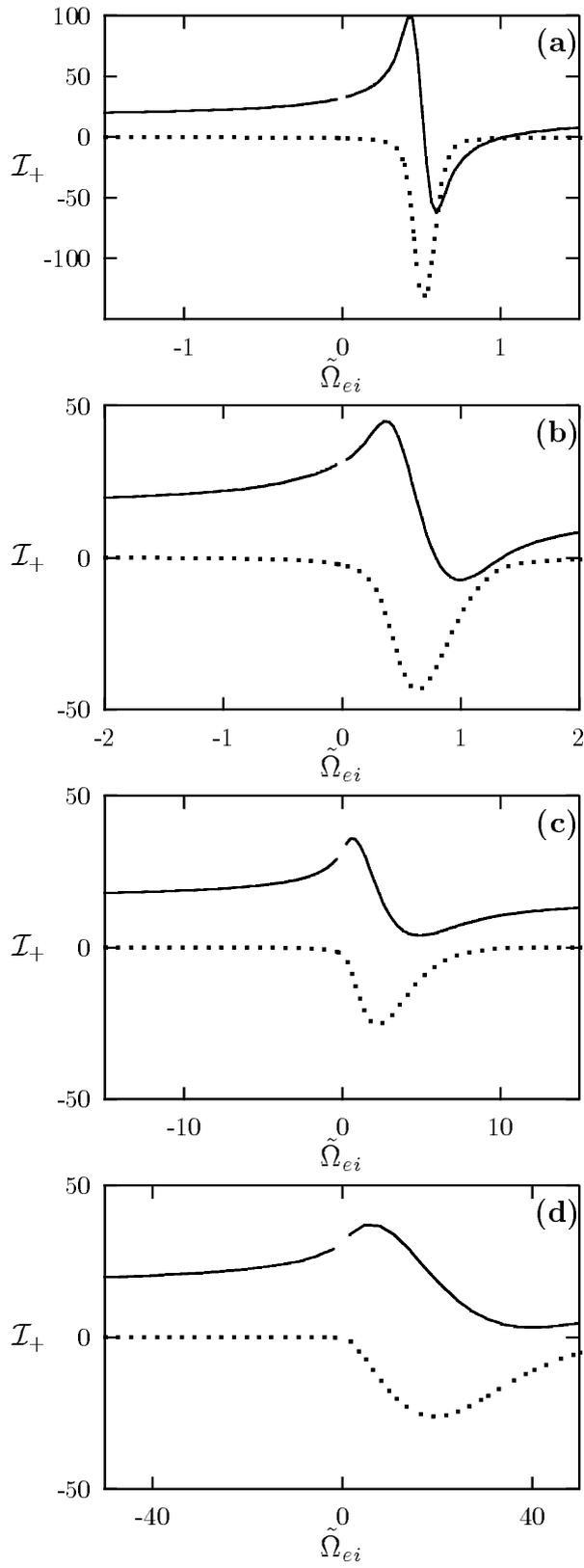
{\small FIG. 2. The real (solid line) and imaginary (dotted line) part of }$%
{\cal I}_{+}\equiv I_{+}\mu \tau _{{\rm ep}}^{2}${\small \ for an
electron-ion plasma at four different temperatures: (a) }$\mu =100${\small ,
(b) }$\mu =10${\small \ ,(c) }$\mu =1${\small \ and (d) }$\mu =0.1${\small .}%
\newline

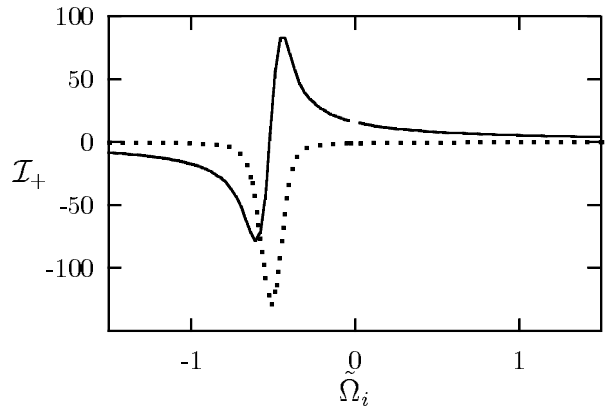
{\small FIG. 3. The real (solid line) and imaginary (dotted line) part of }$%
{\cal I}_{+}\equiv I_{+}\mu \tau _{{\rm ep}}^{2}${\small \ for an
electron-ion plasma for }$\mu _{{\rm e}}=0.1${\small .\ The small region
near }$\tilde{\Omega}_{{\rm i}}=0$ {\small containing the electron
contribution has been left out.}


\begin{references}
\bibitem{Macedo-not}  A term contributing to the force in the Vlasov
equation was incorrectly dropped in Ref. \cite{Macedo} (see the appendix),
which affected the algebraic result somewhat, although not any of the
conclusions made in that paper.
\bibitem{Tetrad}  Eq. (\ref{EFE}) can be transformed to tetrad form simply
by replacing $h_{ab}$ according to $h_{11}=-h_{22}=h_{+}$ and $%
h_{12}=h_{21}=h_{\times }$, and letting \ $\delta T_{ab}$ in (\ref{EFE})
denote tetrad rather than coordinate components.
\bibitem{Numerical note}  Our expression is in exact agreement with some
authors in the limit of zero magnetic field, but otherwise there seem to be
some disagreement over a numerical factor ranging from 8-64 in the
literature, see Ref. \cite{Gayer} p. 1079 for a table of different results.
\end{references}
\end{document}